\documentstyle[epsf]{mn}

\newif\ifAMStwofonts
\AMStwofontstrue

\newcommand{\etal}{et al.}
\newcommand{\eg}{e.g.}
\newcommand{\ie}{i.e.}

\newcommand{\bvtext}{Brunt-V\"ais\"al\"a}

\newcommand{\hetwo}{\hbox{He\,{\sc ii}}}

\newcommand{\bOmega}{\mathbf\Omega}
\newcommand{\omegat}{\omega_{t}}
\newcommand{\omegan}{\omega_{n}}
\newcommand{\llm}{\lambda_{lm}}
\newcommand{\brunt}{N}
\newcommand{\sound}{c_{\rmn s}}
\newcommand{\hough}{\Theta^{m}_{l}}
\newcommand{\legen}{P^{m}_{l}}
\newcommand{\ktr}{k_{\rmn tr}}
\newcommand{\tauacc}{\tau_{\rmn acc}}

\newcommand{\diff}{{\rmn d}}

\newcommand{\teff}{T_{\rmn eff}}
\newcommand{\rsun}{R_{\sun}}
\newcommand{\msun}{M_{\sun}}
\newcommand{\lsun}{L_{\sun}}

\newcommand{\kelv}{\,{\rmn K}}

\ifoldfss
  \newcommand{\rmn}[1] {{\rm #1}}

  \ifCUPmtlplainloaded \else
    \NewTextAlphabet{textbfit} {cmbxti10} {}
    \NewTextAlphabet{textbfss} {cmssbx10} {}
    \NewMathAlphabet{mathbfit} {cmbxti10} {} %
    \NewMathAlphabet{mathbfss} {cmssbx10} {} %
  \fi
  \ifAMStwofonts
    \ifCUPmtlplainloaded \else
      \NewSymbolFont{upmath} {eurm10}
      \NewSymbolFont{AMSa} {msam10}
      \NewMathSymbol{\upi}     {0}{upmath}{19}
      \NewMathSymbol{\umu}     {0}{upmath}{16}
      \NewMathSymbol{\upartial}{0}{upmath}{40}
      \NewMathSymbol{\leqslant}{3}{AMSa}{36}
      \NewMathSymbol{\geqslant}{3}{AMSa}{3E}

       \let\le=\leqslant
       \let\ge=\geqslant
    \fi
  \fi
\fi %

\ifnfssone
  \newmathalphabet{\mathit}
  \addtoversion{normal}{\mathit}{cmr}{m}{it}
  \addtoversion{bold}{\mathit}{cmr}{bx}{it}
  \newcommand{\rmn}[1] {\mathrm{#1}}

  \newmathalphabet{\mathbfit} %
  \addtoversion{normal}{\mathbfit}{cmr}{bx}{it}
  \addtoversion{bold}{\mathbfit}{cmr}{bx}{it}
  \newmathalphabet{\mathbfss} %
  \addtoversion{normal}{\mathbfss}{cmss}{bx}{n}
  \addtoversion{bold}{\mathbfss}{cmss}{bx}{n}
  \ifAMStwofonts
    \ifCUPmtlplainloaded \else
      \UseAMStwoboldmath
      \makeatletter
      \new@mathgroup\upmath@group
      \define@mathgroup\mv@normal\upmath@group{eur}{m}{n}
      \define@mathgroup\mv@bold\upmath@group{eur}{b}{n}
      \edef\UPM{\hexnumber\upmath@group}
      \new@mathgroup\amsa@group
      \define@mathgroup\mv@normal\amsa@group{msa}{m}{n}
      \define@mathgroup\mv@bold\amsa@group{msa}{m}{n}
      \edef\AMSa{\hexnumber\amsa@group}
      \makeatother
      \mathchardef\upi="0\UPM19
      \mathchardef\umu="0\UPM16
      \mathchardef\upartial="0\UPM40
      \mathchardef\leqslant="3\AMSa36
      \mathchardef\geqslant="3\AMSa3E

       \let\le=\leqslant
       \let\ge=\geqslant
    \fi
  \fi
\fi %

\ifnfsstwo
  \newcommand{\rmn}[1] {\mathrm{#1}}

  \DeclareMathAlphabet{\mathbfit}{OT1}{cmr}{bx}{it}
  \SetMathAlphabet\mathbfit{bold}{OT1}{cmr}{bx}{it}
  \DeclareMathAlphabet{\mathbfss}{OT1}{cmss}{bx}{n}
  \SetMathAlphabet\mathbfss{bold}{OT1}{cmss}{bx}{n}
  \ifAMStwofonts
    \ifCUPmtlplainloaded \else
      \DeclareSymbolFont{UPM}{U}{eur}{m}{n}
      \SetSymbolFont{UPM}{bold}{U}{eur}{b}{n}
      \DeclareSymbolFont{AMSa}{U}{msa}{m}{n}
      \DeclareMathSymbol{\upi}{0}{UPM}{"19}
      \DeclareMathSymbol{\umu}{0}{UPM}{"16}
      \DeclareMathSymbol{\upartial}{0}{UPM}{"40}
      \DeclareMathSymbol{\leqslant}{3}{AMSa}{"36}
      \DeclareMathSymbol{\geqslant}{3}{AMSa}{"3E}

       \let\le=\leqslant
       \let\ge=\geqslant
    \fi
  \fi
\fi %

\ifCUPmtlplainloaded \else
  \ifAMStwofonts \else %
    \def\upi{\pi}
    \def\umu{\mu}
    \def\upartial{\partial}
  \fi
\fi

\title[Surface trapping and leakage of $g$-modes in early-type stars --
 I. Qualitative analysis]
 {Surface trapping and leakage of low-frequency $\bmath{g}$-modes in rotating
 early-type stars -- I. Qualitative analysis}
\author[R. H. D. Townsend]
       {R. H. D. Townsend\thanks{E-mail: rhdt@star.ucl.ac.uk}\\
        Department of Physics \& Astronomy, University College London, 
        Gower Street, London WC1E 6BT}
\date{%
Received: .................................... 
Accepted: ....................................
}

\pagerange{\pageref{firstpage}--\pageref{lastpage}}
\pubyear{1999}

\begin{document}

\maketitle

\label{firstpage}

\begin{abstract} 
A qualitative study of the surface trapping of low-frequency
non-radial $g$-modes in rotating early-type stars is undertaken within
the Cowling, adiabatic and traditional approximations. A dispersion
relation describing the local character of waves in a rotating star is
derived; this dispersion relation is then used to construct
propagation diagrams for a $7\,\msun$ stellar model, which show the
location and extent of wave trapping zones inside the star. It is
demonstrated that, at frequencies below a cut-off, waves cannot be
fully trapped within the star, and will leak through the
surface. Expressions for the cut-off frequency are derived in both the
non-rotating and rotating cases; it is found from these expressions
that the cut-off frequency increases with the rotation rate for all
but prograde sectoral modes.

While waves below the cut-off cannot be reflected at the stellar
surface, the presence of a sub-surface convective region in the
stellar model, due to \hetwo\ ionization, means that they can become
partially trapped within the star. The energy leakage associated with
such waves, which are assigned the moniker {\em virtual modes} due to
their discrete eigenfrequencies, means that stability analyses which
disregard their existence (by assuming perfect reflection at the
stellar surface) may be in error.

The results are of possible relevance to the 53 Per and SPB classes of
variable star, which exhibit pulsation frequencies of the same order
of magnitude as the cut-off frequencies found for the stellar
model. It is suggested that observations either of an upper limit on
variability periods (corresponding to the cut-off), or of line-profile
variations due to virtual modes, may permit asteroseismological
studies of the outer layers of these systems.
\end{abstract}


\section{Introduction} \label{sec:introduction}

The self-excitation of global non-radial pulsation modes in a star is
a prime example of positive feedback, whereby small oscillatory
perturbations grow in amplitude via the efficient conversion of heat
into vibrational energy by a suitable driving mechanism (see, \eg,
Unno \etal\ 1989 for a comprehensive review of the topic). A
fundamental ingredient in the feedback loop is that the oscillations
must be trapped in some part of the stellar interior, so that energy
does not leak from the system faster than it can be generated. Such
trapping can occur when a pair of evanescent regions, where traveling
waves cannot be supported, enclose a propagative region; waves are
repeatedly reflected at the two evanescent boundaries, and the
resulting superposition leads to a standing wave of the normal-mode
type.

For waves excited in stellar envelopes, it is common for the surface
layers to serve as one of the evanescent regions required for the
formation of a trapping zone. Ando \& Osaki \shortcite{AndOsa1975}
demonstrated that such a situation occurs in the Sun, where low-order
$p$-modes are trapped beneath the photosphere, supporting a model
first put forward by Ulrich \shortcite{Ulr1970} to explain the
five-minute solar oscillation \cite{LeiNoySim1962}. However, the
trapping is only effective for modes with frequencies below some
cut-off; higher-frequency modes cannot be reflected at the
photosphere, and will leak through the stellar surface. This issue was
addressed in detail by Ando \& Osaki \shortcite{AndOsa1977}, who found
that, although leakage does occur through the solar photosphere at
frequencies above the cut-off, some waves can subsequently be
reflected at the chromosphere-corona interface, and standing waves are
able to form. More recently, Balmforth \& Gough \shortcite{BalGou1990}
suggested that such coronal reflection can explain apparent
observations of high-frequency chromospheric standing waves
\cite{FleDeu1989}, although debate concerning this interpretation
still continues \cite{Kum1994,DzhSta1995,Jef1998}.

Pulsation in massive, early stars (types O and B) is qualitatively
quite different from the solar case, due to the gross structural
differences between the two stellar classes. However, it is still
subject to the same wave trapping requirements, since the underlying
physics remains the same. Shibahashi \& Osaki \shortcite{ShiOsa1976},
in their study of $g$-modes trapped within the hydrogen-burning shell
of evolved massive stars, found that high-frequency (low-order) modes
can tunnel through an evanescent region separating core and envelope,
and thence escape from the star.  A complementary situation was
discussed by Osaki \shortcite{Osa1977} when studying pulsation in
Cepheid-type stars; non-radial $p$-modes trapped within the envelope
were able to tunnel through an evanescent region into the core, where
they were damped rapidly without reflection at the centre.

In both cases, the appropriate region of the star was modeled as an
isolating oscillating unit with the inclusion of wave leakage at one
boundary. The leakage was found to stabilize some modes which would
otherwise have been self-excited, due to the associated loss of
vibrational energy from the star. Shibahashi \shortcite{Shi1979}
analyzed wave trapping in an idealized stellar model (corresponding to
an evolved massive star) using an asymptotic method, and discussed in
some depth these two cases; in addition, he considered the situation
where low-frequency (high-order) $g$-modes are able to tunnel through
an evanescent region in the envelope and thence escape through the
stellar surface.

More recently, however, relatively little attention has been shown
regarding wave trapping issues at the surface of early-type stars; in
particular, stability analyses
\cite{Cox1992,Kir1992,DziPam1993,GauSai1993}, based on the new
opacity calculations of Rogers \& Iglesias \shortcite{RogIgl1992} and
Seaton \shortcite{Sea1992}, have assumed that the Lagrangian pressure
perturbation $\delta p$ tends to zero or some limiting value at the
stellar surface. Such an assumption corresponds to the {\em ab initio}
condition that waves incident from the interior are totally reflected
at the stellar surface; the possibility of leakage is thereby
disregarded, and no consideration of trapping issues is undertaken.

This is the first in a short series of papers studying the surface
trapping of low-frequency $g$-modes in early-type stars, in an attempt
to re-open discussion of, and investigation into, this important
area. Much of the work is conceptually developed from that of Ando \&
Osaki \shortcite{AndOsa1975}; however, in light of recent research
into the influence of rotation on low-frequency modes
\cite{LeeSai1990,Bil1996,LeeSai1997}, and due to the fact that
significant rotation appears to be commonplace in O- and B-star
populations \cite{How1997}, the theory is updated to include
rotational effects.

The current paper serves as a introduction, covering the more
qualitative, general aspects of the study; subsequent papers will
investigate various issues arising from this paper in greater
depth. The following section reviews the pulsation equations
appropriate for low-frequency $g$-modes in rotating stars, whilst
section \ref{sec:dispersion} derives the dispersion relation
corresponding to these equations. The trapping of waves described by
this dispersion relation is examined in section \ref{sec:trapping}
with the aid of propagation diagrams, and the effect of rotation on
the eigenfrequencies of individual modes is discussed in
section \ref{sec:eigenfrequency}. The findings are discussed in
section \ref{sec:discussion}, and summarized in
section \ref{sec:conclusions}.

\section{Pulsation equations} \label{sec:pulsation}

The dynamics of pulsation in a rotating star differ from the
non-rotating case due to the influence of the fictitious Coriolis and
centrifugal forces, which arise as a consequence of the non-inertial
nature of a rotating frame of reference. The centrifugal force breaks
the equilibrium symmetry of the star, so that the level
(equipotential) surfaces become oblate spheroids rather than the usual
concentric spheres. Such a change in stellar configuration will
manifest itself implicitly in the pulsation equations, through
modifications to the equilibrium variables of state. However, in the
case of uniform (solid-body) rotation, no {\em explicit} modification
of the pulsation equations occurs due to this centrifugal distortion
\cite{Unn1989}. In contrast, the Coriolis force enters the pulsation equations
explicitly, through the introduction of a velocity-dependent term in
the hydrodynamical momentum equation. This term can lead to the
significant modification of individual pulsation modes, and is also
responsible for the existence of new classes of wave-like solutions
\cite{Lon1968} which are not found in non-rotating systems.

Simultaneous treatment of both forces within a pulsation framework is
fraught with difficulty. Some progress towards this goal has been made
(\eg, Lee 1993; Lee \& Baraffe 1995), but attempts remain frustrated
by the fact that the centrifugal distortion cannot really be
considered as an {\em a posteriori} modification to the structure of a
given star, but must be treated self-consistently with the evolution
of the star (see, \eg, Meynet \& Maeder 1997).  However, in a number
of limiting cases, certain approximations can be made which simplify
the problem significantly. In the case of the low-frequency modes, the
Coriolis force will dominate the centrifugal force, and the effects of
the latter on the equilibrium configuration may be disregarded if the
rotation is not too severe. Furthermore, the so-called `traditional
approximation'
\cite{Eck1960} may be employed, whereby the horizontal component of
the angular frequency vector of rotation $\bOmega$ is neglected. This
approximation is most appropriate for low-frequency pulsation modes in
the outer regions of a star \cite{Unn1989}, and therefore can be
considered useful in the present study.

In combination with the Cowling \shortcite{Cow1941} and adiabatic
approximations, where the perturbations to the gravitational potential
and specific entropy, respectively, are neglected, the traditional
approximation renders the pulsation equations separable in the
spherical polar co-ordinates $(r,\theta,\phi)$. Solutions for the
dependent variables $\xi_{r}$ and $p'$, the radial fluid displacement
and Eulerian pressure perturbation, respectively, may then be written
in the form \cite{LeeSai1997}
\begin{equation} \label{eqn:solution1}
\xi_{r} = \xi_{r}(r) \hough(\mu;\nu) \exp[{\rmn i}(m \phi + \omega t)],
\end{equation}
\begin{equation} \label{eqn:solution2}
p' = p'(r) \hough(\mu;\nu) \exp[{{\rmn i}(m \phi + \omega t)}],
\end{equation}
where $\mu \equiv \cos \theta$ is the normalized latitudinal distance
from the equatorial plane, $\omega$ is the pulsation frequency in the
co-rotating reference frame, and $\hough(\mu;\nu)$ is a Hough function
\cite{Bil1996,LeeSai1997}. These Hough functions are the
eigensolutions of Laplace's tidal equation \cite{Lon1968}, and form a
one-parameter family in $\nu \equiv 2\Omega/\omega$, where $\Omega
\equiv |\bOmega|$ is the angular frequency of rotation. The integer
indices $l$ and $m$, with $l\ge0$ and $|m|\le l$, correspond to the
harmonic degree and azimuthal order, respectively, of the associated
Legendre polynomials $\legen(\mu)$ \cite{AbrSte1964} to which the
Hough functions reduce in the non-rotating limit, so that $\hough(\mu;
0) \equiv \legen(\mu)$. This indexing scheme, based on the one adopted
by Lee \& Saio \shortcite{LeeSai1990}, is less general than that of
Lee \& Saio \shortcite{LeeSai1997}, in that it does not encompass the
Hough functions corresponding to Rossby and oscillatory convective
modes (which do not have non-rotating counterparts); however, such
modes are not considered herein, and the current scheme is
sufficient. Note that $\omega$ is considered to be positive throughout
the following discussion, and, therefore, prograde and retrograde
modes correspond to negative and positive values of $m$, respectively.

The radial dependence of the solutions
(\ref{eqn:solution1}--\ref{eqn:solution2}) is described by the
eigenfunctions $\xi_r(r)$ and $p'(r)$, which are governed by a pair of
coupled first-order differential equations. In order to facilitate
subsequent manipulation, it is useful to write these equations in the
form
\begin{equation} \label{eqn:pulsation1}
\frac{1}{r^{2}}\frac{\diff}{\diff r}\left(r^{2}\xi_{r}\right) -
\frac{g}{\sound^{2}}\xi_{r} =
\frac{1}{\omega^{2}\sound^{2}}
\left(
 \frac{\llm^{2}\sound^{2}}{r^{2}} - \omega^{2}
\right)
\frac{p'}{\rho}
\end{equation}
and
\begin{equation} \label{eqn:pulsation2}
\frac{1}{\rho}\frac{\diff p'}{\diff r} +
\frac{g}{\sound^{2}} \frac{p'}{\rho} =
(\omega^{2} - N^{2}) \xi_r,
\end{equation}
where $\rho$, $\sound$, $g$ and $\brunt$ are the local equilibrium
values of the density, adiabatic sound speed, gravitational
acceleration and \bvtext\ frequency, respectively. Note that $\xi_{r}$
and $p'$ are now taken to be functions of $r$ alone in both these and
subsequent equations, unless explicitly stated.

The quantity $\llm$ appearing in equation (\ref{eqn:pulsation1}),
which arises as separation constant when solutions of the form
\mbox{(\ref{eqn:solution1}--\ref{eqn:solution2})} are sought, is the
eigenvalue of Laplace's tidal equation corresponding to the
appropriate Hough function $\hough(\mu;\nu)$. In the limit $\nu = 0$,
this eigenvalue is equal to $l(l+1)$, and equations
(\ref{eqn:pulsation1}--\ref{eqn:pulsation2}) are then identical to
those appropriate for a non-rotating star (\eg, Unno \etal\ 1989,
\S15.1).  The utility of the traditional approximation thus lies in
the fact that much of the formalism of the non-rotating case may also
be applied to rotating stars with the simple replacement of $l(l+1)$
by $\llm$, a result first found by Lee \& Saio \shortcite{LeeSai1987a}.

Global solution of equations
(\ref{eqn:pulsation1}--\ref{eqn:pulsation2}) must typically be
approached numerically; however, an examination of the local character
of the solutions suffices in the present qualitative context. This
character is governed by the dispersion relation applicable to the
equations, discussed in the following section.

\section{Dispersion relation} \label{sec:dispersion}

To derive a local dispersion relation for the pulsation equations
(\ref{eqn:pulsation1}--\ref{eqn:pulsation2}), it is useful first to
place the equations in a canonical form similar to that introduced by
Osaki \shortcite{Osa1975} for the non-rotating case. By defining the
two new eigenfunctions,
\begin{equation}
\tilde{\xi} = r^{2}\xi_{r}\exp
 \left(
   -\int_{0}^{r}\frac{g}{\sound^{2}}\, \diff r
 \right),
\end{equation}
\begin{equation}
\tilde{\eta} = \frac{p'}{\rho}\exp
 \left(
   -\int_{0}^{r}\frac{N^{2}}{g}\, \diff r
 \right),
\end{equation}
the left-hand sides of both pulsation equations may be written as a
single derivative, and the canonical form is found as
\begin{equation}
\frac{\diff \tilde{\xi}}{\diff r} = h(r)
\frac{r^{2}}{\sound^{2}\omega^{2}}
 \left(
  \frac{\llm \sound^{2}}{r^{2}} - \omega^{2}
 \right) \tilde{\eta},
\end{equation}
\begin{equation}
\frac{\diff \tilde{\eta}}{\diff r} = \frac{1}{r^{2}h(r)}
 \left(
   \omega^{2} - \brunt^{2}
 \right) \tilde{\xi},
\end{equation}
where
\begin{equation}
h(r) = \exp \left[ \int_{0}^{r} 
  \left( \frac{N^{2}}{g} - \frac{g}{\sound^{2}} \right )
 \,\diff r \right].
\end{equation}
Note that $h(r)$ is always positive, so that the original
eigenfunctions $\xi_{r}$ and $p'$ everywhere share the same sign as
$\tilde{\xi}$ and $\tilde{\eta}$, respectively.

Qualitative solution of these canonical equations is accomplished
using the same method as Osaki \shortcite{Osa1975}, namely, by
assuming that the coefficients on the right-hand sides are independent
of $r$. Such an assumption will be valid if the characteristic
variation scale of the solutions is much smaller than that of the
coefficients. Then, local solutions of the form
\begin{equation} \label{eqn:localsolutions}
\tilde{\xi}, \tilde{\eta} \sim \exp({\rmn i}k_{r} r),
\end{equation}
lead to a dispersion relation for the radial wavenumber
$k_{r}$, 
\begin{equation} \label{eqn:dispersion1}
k_{r}^{2} = \frac{1}{\sound^{2}\omega^{2}}
\left(
 \frac{\llm\sound^{2}}{r^{2}} - \omega^{2}
\right)
\left(
 \brunt^{2} - \omega^{2}
\right).
\end{equation}
By introducing the effective transverse wavenumber $\ktr$, defined by
Bildsten \etal\ \shortcite{Bil1996} as
\begin{equation} \label{eqn:transverse1}
\ktr^{2} = \frac{\llm}{r^{2}},
\end{equation}
the dispersion relation may be re-written in the more useful form
\begin{equation} \label{eqn:dispersion2}
k_{r}^{2}\sound^{2}\omega^{2} =
\left(\ktr^{2} \sound^{2} - \omega^{2}\right)
\left(N^{2} - \omega^{2}\right),
\end{equation}
The value of $k_{r}$ for given $\omega$ and $r$, calculated using this
expression, determines the local character of waves at the appropriate
frequency and location within the star. Inspection of equation
(\ref{eqn:localsolutions}) shows that real values ($k_{r}^{2} > 0$)
correspond to propagative regions, where the waves oscillate
spatially, whilst imaginary values ($k_{r}^{2} < 0$) correspond to
evanescent regions, where the waves grow or decay exponentially in
amplitude. The $k_{r}^{2} = 0$ curves in the $(r,\omega^{2})$ plane,
defined by the roots of the right-hand side of the dispersion
relation, separate these two types of region, and therefore correspond
to the reflective boundaries discussed in the introduction. These
boundaries, of fundamental importance when trapping zones are
considered, are examined in the following section with the aid of
propagation diagrams. Note that this $k_{r}^{2} = 0$ definition of the
reflective boundaries formally violates the assumption used previously
to derive the solutions (\ref{eqn:localsolutions}); however, this
violation will have little effect on the positions of the boundaries,
and is not important at a qualitative level.

\begin{table}
\centering
\caption{The physical parameters of the stellar model considered throughout.}
\label{tab:model}
\begin{tabular*}{80mm}{@{\extracolsep{\fill}}ccccc} 
\hline
$M/\msun$ & $R/\rsun$ & $L/\lsun$ & $\teff/\kelv$ & $Z$ \\ 7.00 & 3.16
& $1.76 \times 10^{3}$ & 21,000 & 0.02 \\
\hline
\end{tabular*}
\end{table}

The remainder of this section is left to a discussion of the effective
transverse wavenumber $\ktr$, since, as will be demonstrated
subsequently, this quantity can be pivotal in determining the trapping
conditions at the stellar surface. In the case of plane waves in an
infinite, plane-parallel, stratified medium, $\ktr$ may be regarded as
a free parameter; however, in the case of a spherical configuration,
its is constrained to assume values permitted by equation
(\ref{eqn:transverse1}).  These constraints arise from to transverse
boundary conditions applicable to waves propagating in horizontal
(\ie, non-radial) directions; in a non-rotating star, they are
equivalent to the requirement that solutions are invariant under the
periodic transformations $\theta \rightarrow \theta+2\pi$ and $\phi
\rightarrow \phi+2\pi$, and lead to the familiar result (\eg, Unno
\etal\ 1989)
\begin{equation} \label{eqn:transverse2}
\ktr^{2} =
\frac{l(l+1)}{r^{2}} \qquad (\nu = 0).
\end{equation} 

When significant rotation is introduced, both of these requirements
still hold, but an additional constraint in $\theta$ is introduced as
a consequence of the variation of the Coriolis force with
latitude. Such variation means that, for $\nu > 1$, waves near the
equator which are propagating in the latitudinal direction become
evanescent when $|\mu|> 1/\nu$; subsequent reflections lead to the
trapping of these waves within the so-called `equatorial waveguide'
\cite{Gil1982}. The resulting horizontally-standing waves, whose
angular dependence is described by the Hough functions
$\hough(\mu;\nu)$, are oscillatory in latitude between the waveguide
boundaries at $\mu = \pm 1/\nu$, and evanescent elsewhere.  With
increasing $\nu$, these boundaries converge towards the equator; for
significant rotation, the constraints on $\ktr$ therefore become
dominated by the approximate requirement that an integer number of
half-wavelengths in latitude fit between the waveguide
boundaries. This requirement is manifest in the asymptotic expression
for $\llm$ found by Bildsten \etal\ \shortcite{Bil1996}, which, when
substituted into equation (\ref{eqn:transverse1}), gives
\begin{equation} \label{eqn:transverse3}
\ktr^{2} = \frac{(2l_{\mu}-1)^{2}\nu^{2}}{r^{2}} \qquad (\nu \gg 1),
\end{equation}
where $l_{\mu}$ is the number of latitudinal nodes exhibited by the
appropriate Hough function between the waveguide boundaries. This
latter quantity is independent of $\nu$ for prograde ($m>0$) and zonal
($m=0$) modes, whilst it increments by 2 as $\nu$ is increased beyond
unity for retrograde ($m<0$) modes, due to the introduction of an
additional pair of latitudinal nodes at $\nu = 1$
\cite{LeeSai1990,LeeSai1997}. Furthermore, $l_{\mu} = l - |m|$ for $\nu
= 0$, since the associated Legendre polynomials $\legen(\mu)$ exhibit
$l-|m|$ zeroes over $-1 < \mu < 1$, and $\hough(\mu;0) \equiv
\legen(\mu)$. These properties mean that $l_{\mu}$ in the above
asymptotic expression (\ref{eqn:transverse3}) may be written in terms
of $l$ and $m$ as
\begin{equation} \label{eqn:transverse4}
\ktr^{2} = 
\frac{(2l-2|m|\pm 1)^{2}\nu^{2}}{r^{2}} \qquad (\nu \gg 1),
\end{equation}
the plus sign being chosen for retrograde modes ($m>0$), and the minus
sign for prograde and zonal modes ($m \le 0$). A comparison of this
result with equation (\ref{eqn:transverse2}) indicates that the
permitted values of $\ktr$ for $\nu \gg 1$ can greatly exceed the
corresponding ones in the non-rotating case, especially for small
$|m|$.

The notable exception to this discussion is the case of the prograde
sectoral modes ($m = -l$), which in the limit $\nu \gg 1$ are
transformed into equatorially-trapped Kelvin waves. Such Kelvin waves
have a exponential latitudinal dependence at small $\mu$ described by
\cite{Gil1982}
\begin{equation}
\xi_{r}, p' \sim \exp \left(-\frac{\Omega^{2}\mu^{2}r^{2}}{\sound^{2}}\right),
\end{equation}
indicating that they should be considered evanescent in the
latitudinal direction even at the equator. Therefore, the constraints
on $\ktr$ are dominated by the periodic boundary condition in $\phi$;
the $\exp ({\rmn i}m\phi)$ azimuthal dependence of the solutions
(\ref{eqn:solution1}--\ref{eqn:solution2}) then gives the transverse
wavenumber for prograde sectoral modes as
\begin{equation} \label{eqn:transverse5}
\ktr^{2} = \frac{m^{2}}{r^{2}} \qquad (\nu \gg 1, m = -l),
\end{equation}
which can also be derived using the asymptotic expression for $\llm$
found by Bildsten \etal\ \shortcite{Bil1996} for these modes.

\begin{figure}
\epsffile{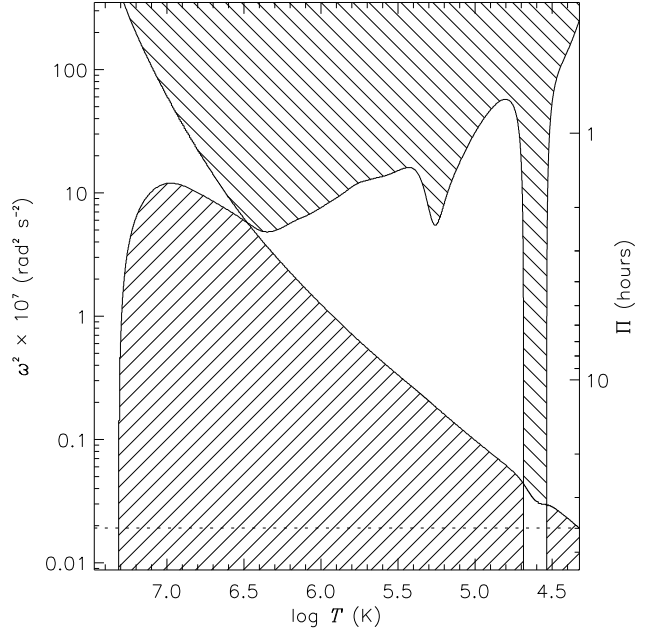}
\caption{The propagation diagram for $l=4$, $m=-1$ modes in the $7\,\msun$
stellar model, plotted in the $(\log T, \omega^2)$ plane to emphasize
the outer regions of the model; the period $\Pi$ corresponding to the
frequency is shown on the right-hand ordinate. Hatched areas
correspond to propagative regions where waves have $g$-mode ($/$) or
$p$-mode ($\backslash$) character, whilst evanescent regions are
blank. The dotted horizontal line shows the position of the trapping
cut-off frequency $\omegat$.}
\label{fig:propdiag1}
\end{figure}

\section{Wave trapping and leakage} \label{sec:trapping}

As was demonstrated in the preceding section, the character of waves
within a star is determined by the local radial wavenumber, so that
positive and negative values of $k_{r}^{2}$ can be identified with
propagative and evanescent regions, respectively. An indispensable
diagnostic tool for visualizing the location and extent of these
regions, over a range of frequencies, is the propagation diagram
introduced by Scuflaire~\shortcite{Scu1974}, in which the $(r,
\omega^{2})$ plane is divided into zones over which the sign of
$k_{r}^{2}$ is constant.

Figure \ref{fig:propdiag1} shows the propagation diagram for $l=4$
modes in a typical (non-rotating) early-type star; the logarithm of
the temperature $T$ has been adopted as the abscissa, rather than the
radius, to emphasize the outer regions of the star. Regions in the
$(\log T, \omega^{2})$ plane where waves are propagative ($k_{r}^{2} >
0$) are hatched, whilst evanescent regions ($k_{r}^{2} < 0$) are
blank. Values for the \bvtext\ frequency $\brunt$ and adiabatic sound
speed $\sound$ throughout the star, required for the evaluating
$k_{r}^{2}$ using the dispersion relation (\ref{eqn:dispersion2}),
have been taken from a $7\,\msun$ ZAMS stellar model, calculated by
Loeffler \shortcite{Loe1998}, whose parameters summarized in table
\ref{tab:model}; the model extends out to the photosphere at optical
depth $\tau = 2/3$, where the temperature $T \equiv \teff=21,000\kelv$
corresponds to an early-B spectral type. Equation
(\ref{eqn:transverse2}), which is appropriate in the non-rotating
case, has been used to calculate $\ktr$.

\begin{figure}
\epsffile{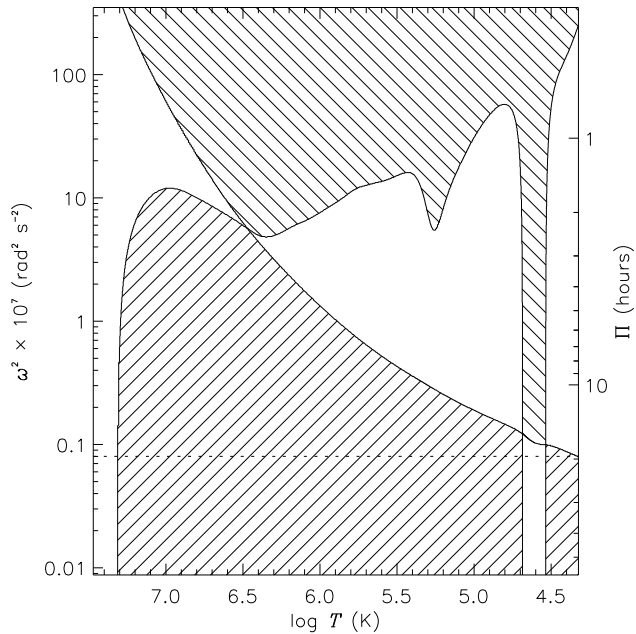}
\caption{As for fig.\ref{fig:propdiag1}, but rotation has been
introduced at an angular frequency $\Omega = 8.04 \times
10^{-5}\,{\rmn rad\ s^{-1}}$. The azimuthal order $m$ is -1.}
\label{fig:propdiag2}
\end{figure}

In this figure, the type of hatching used to show propagative regions
delineates between waves with $p$- and $g$-mode characters, the former
occurring when both parenthetical terms on the right hand side of the
dispersion relation (\ref{eqn:dispersion2}) are negative, and the
latter when both terms are positive. The division of the diagram into
relatively distinct $p$- and $g$-mode propagation regions is
characteristic of early-type stars. This division arises due to the
fact that $\ktr^{2}\sound^{2}$ diverges at the origin (due to the
$1/r$ dependence of $\ktr$) and is relatively small at the surface,
while $\brunt^{2}$ is approximately zero in the convective core ($\log
T \ga 7.3$) and relatively large at the surface due to the steep
stratification there. The prominent `well' at $\log T \approx 4.6$
indicates the presence of a thin convective region ($\brunt^{2} < 0$)
due to \hetwo\ ionization, whilst the smaller well at $\log T \approx
5.3$ is due to the metal opacity bump responsible for
$\kappa$-mechanism pulsation in early-type stars.

As mentioned in the previous section, the $k_{r}^{2}=0$ curves which
separate propagative and evanescent regions correspond to the
reflective boundaries required for wave trapping. Inspection of
fig. \ref{fig:propdiag1} shows that, for $g$-modes with frequencies
greater than the trapping cut-off $\omegat$, where $\omegat^{2}
\approx 1.9 \times 10^{-9}\,{\rmn rad^{2}\ s^{-2}}$ is shown in the
figure as a horizontal dotted line, there exists an extensive trapping
zone formed by a pair of reflective boundaries, one at the edge of the
convective core and the other in the envelope at lower
temperatures. In contrast, for modes with frequencies below $\omegat$,
waves are propagative even at the surface of the star, and the outer
reflective boundary required for the formation of a trapping zone does
not exist.

Strictly speaking, the term `mode' is not appropriate in such
circumstances, since stationary waves will not be established by
repeated complete reflection. However, all waves at frequencies below
the cut-off are evanescent in the convective region at $\log T \approx
4.6$. This region, with a width of approximately 0.16 percent of the
stellar radius, behaves like a partially-reflecting barrier to waves
incident from the interior; some fraction of the waves will leak
through the barrier and thence propagate unhindered to the surface,
where they are lost from the star, whilst the remaining reflected
fraction will contribute to the establishment of `somewhat-stationary'
waves interior to the barrier. Within the adiabatic approximation,
these waves must decay exponentially in amplitude with time to
compensate for the the energy lost through leakage, but will still
exhibit a discrete eigenfrequency spectrum. Shibahashi \& Osaki
\shortcite{ShiOsa1976}, when considering a similar situation for
high-frequency $g$-modes in evolved early-type stars, drew a useful
analogy with virtual levels in the potential problem of quantum
mechanics; therefore, it seems appropriate to refer to such
partially-trapped waves as {\em virtual modes}. Whether virtual modes
can actually be self-excited in a star depends on the balance between
the input of vibrational energy from a suitable driving mechanism, and
the loss of vibrational energy associated with the leakage;
non-adiabatic calculations are required to answer such a questions.

The trapping cut-off frequency $\omegat$, which separates the leaking
virtual modes from the fully-trapped `traditional' modes, is given by
the smaller root of the dispersion relation (\ref{eqn:dispersion2}) at
the stellar surface, namely
\begin{equation} \label{eqn:trapping1}
\omegat^{2} = \left.\ktr^{2}\sound^{2}\right|_{r=R},
\end{equation}
where $R$ is the stellar radius. This expression demonstrates the
pivotal r\^{o}le of the effective transverse wavenumber $\ktr$,
discussed at the end of the preceding section, in determining the
trapping condition at the surface. In the non-rotating context, $\ktr$
can be eliminated from this expression through use of equation
(\ref{eqn:transverse2}) to give
\begin{equation} \label{eqn:trapping2}
\omegat^{2} = \left.
\frac{l(l+1)\sound^{2}}{r^{2}}\right|_{r=R}
\end{equation}
for $\Omega = 0$.

When the effects of rotation are included, the more general expression
(\ref{eqn:transverse1}) for $\ktr$ must be used in evaluating the sign
of $k_{r}^{2}$ using the dispersion relation
(\ref{eqn:dispersion2}). However, propagation diagrams may be
constructed and interpreted in exactly the same manner as the
non-rotating case. Figure \ref{fig:propdiag2} shows the propagation
diagram for the $7\,\msun$ stellar model considered previously, but
with rotation included at an angular frequency $\Omega = 8.04 \times
10^{-5}\,{\rmn rad\ s^{-1}}$, which is half of the critical rotation
rate for the star, and corresponds to a period of 21.7 hours. The
effects of the rotation on the equilibrium stellar structure having
been neglected. Calculation of the eigenvalue $\llm$ in equation
(\ref{eqn:transverse1}), for each frequency ordinate value in the
$(\log T, \omega^{2})$ plane, was accomplished using Townsend's
\shortcite{Tow1997} implementation of the matrix eigenvalue method
presented by Lee \& Saio \shortcite{LeeSai1990}. This method
corresponds to the spectral expansion of Hough functions in a
truncated series of associated Legendre polynomials of the same
azimuthal order $m$; 100 expansion terms were used throughout the
calculations, a value deemed to provide sufficient accuracy since a
similar calculation with 200 terms produced no numerical change in the
results. An azimuthal order $m=-1$ was adopted, so
fig. \ref{fig:propdiag2} should be taken as appropriate for modes with
$(l,m)=(4,-1)$.

Inspection of this figure shows that the trapping cut-off is
significantly larger ($\omegat^{2} \approx 8.0 \times 10^{-9}\,{\rmn
rad^{2}\ s^{-2}}$) than in the non-rotating case. This is a direct
consequence of the influence of rotation on $\ktr$; at low frequencies
where $\omega < 2\Omega$, $\nu > 1$, and $\ktr$ can assume large
values, as discussed in the preceding section. The appropriate
expression for $\omegat$ in rotating stars is given by
\begin{equation} \label{eqn:trapping3}
\omegat^{2} = \left.\frac{\llm\sound^{2}}{r^{2}}\right|_{r=R},
\end{equation}
although this should be regarded as formal, since it must be
remembered that $\llm$ is itself a function of $\omega$ through its
dependence on the parameter $\nu$. However, in the limit $\Omega^{2}
\gg \sound^{2}/r^{2}$ (at the surface), this expression will have
solutions corresponding to $\nu \gg 1$, and thus the asymptotic
expressions (\ref{eqn:transverse4},\ref{eqn:transverse5}) found
previously may be used in the place of the general expression
(\ref{eqn:trapping1}) for $\ktr$. Solving the resulting equations for
$\omegat$ then gives
\begin{equation} \label{eqn:trapping4}
\omegat^{2} = \left\{
\begin{array}{l}
2\Omega (2l-2|m|+1) \sound/r \\
2\Omega (2l-2|m|-1) \sound/r \\
m^{2} \sound^{2}/r^{2}
\end{array}
\right|_{r=R}
\begin{array}{l}
(m> 0) \\
(-l < m \le 0) \\
(m = -l)
\end{array}
\end{equation}
for $\Omega^{2} \gg \sound^{2}/r^{2}$. Applying the middle expression
to the $7\,\msun$ model for $\Omega = 8.04 \times 10^{-5}\,{\rmn rad\
s^{-1}}$, and $\sound/r = 9.76 \times 10^{-6}\,{\rmn s^{-1}}$ at the
surface, leads to the asymptotic value $\omegat^{2} = 7.85 \times
10^{-9}\,{\rmn rad^{2}\ s^{-2}}$ for $(l,m)=(4,-1)$ modes, which is in
reasonably good agreement with the value $\omegat^{2} \approx 8.0
\times 10^{-9} {\rmn rad^{2}\ s^{-2}}$ shown in fig. \ref{fig:propdiag2}.

The above expressions, when compared with equation
(\ref{eqn:trapping2}), demonstrate that the effect of rotation is to
increase the trapping cut-off $\omegat$ for all but the prograde
sectoral modes; these latter modes will exhibit a smaller cut-off in
rotating stars than in the non-rotating case, due to their
transformation into Kelvin waves discussed previously. This result is
interesting in light of anecdotal observational evidence favouring
prograde sectoral modes as the source of periodic line-profile
variations in rapidly-rotating early-type stars. If such evidence can
be substantiated at a quantitative level, as has been done by Howarth
\etal\ \shortcite{How1998} for the rapidly-rotating pulsators HD\,93521
and HD\,64760, then it can be suggested that the bias towards prograde
sectoral modes is due to the suppression of other types of mode, which
will have a large values of $\omegat$ at rapid rotation rates and
therefore preferentially leak from the star without self-excitation.

\section{Eigenfrequencies} \label{sec:eigenfrequency}

In addition to its influence on the trapping cut-off frequency
$\omegat$, rotation modifies the eigenfrequencies and eigenfunctions
of individual modes through its influence on the positions of trapping
boundaries; this can be anticipated from the appearance of $\llm$ in
the pulsation equations
(\ref{eqn:pulsation1}--\ref{eqn:pulsation2}). To evaluate the modified
eigenfrequencies at a qualitative level, the asymptotic technique
developed by Shibahashi \shortcite{Shi1979} and Tassoul
\shortcite{Tas1980} may be adapted using the traditional approximation
to given expressions appropriate for rotating stars. Using such an
approach, Lee \& Saio \shortcite{LeeSai1987a} found that low-frequency
$g$-modes trapped between the boundary of the convective core
($r=r_{c}$) and the surface of a rotating early-type star have
eigenfrequencies $\omegan$ given by
\begin{equation} \label{eqn:frequency}
\omegan = \frac{2\sqrt{\llm}}{(n+\eta_{e}/2-1/6)\pi} \int_{r_{c}}^{R}
\frac{|N|}{r}\,\diff r,
\end{equation}
where $\eta_{e}$ is the effective polytropic index at the surface, and
$n$ is the radial order of the mode. This expression is not strictly
appropriate in the current context, due to the fact that the outer
reflecting boundary for trapped modes in figs.  \ref{fig:propdiag1}
and \ref{fig:propdiag2} occurs at $r<R$; furthermore, the presence of
the convective region at $\log T \approx 4.6$ has been
neglected. However, the form of the expression demonstrates that the
frequencies of individual modes share the same $\llm$-dependence as
the trapping cut-off $\omegat$ in equation (\ref{eqn:trapping3}.

It can therefore be suggested that a $g$-mode which is trapped in a
non-rotating star will remain trapped once rotation is introduced,
since the effect of rotation is to scale both sides of the trapping
condition $\omegan > \omegat$ by an equal amount. Disregarding the
possibility of avoided crossings (Aizenman, Smeyers \& Weigert 1977;
Lee \& Saio 1989), a more general hypothesis may be put forward that,
{\em ceteris paribus}, the set of radial orders $\{n\}$ of the
$g$-modes which are trapped in a non-rotating star will remain
invariant under the influence of rotation, with a similar result
applying to the virtual modes. The hypothesis can be supported with an
analogy drawn to atomic energy levels under the influence of a
magnetic field; even though the levels are distorted by the action of
the Lorentz force (which, like the Coriolis force, can be expressed as
a velocity cross-product), the set of discrete states which are bound
is invariant under the action of the field. However, numerical
calculations should be employed to test this hypothesis rigorously.

\section{Discussion} \label{sec:discussion}

An important caveat regarding {\em quantitative} interpretation of the
results presented previously is that atmospheric layers above the
photosphere at $\tau = 2/3$ have been disregarded. This is justifiable
if $\ktr$ is constant throughout these layers; however, such a
situation is unlikely to be realized, since even in the case of
isothermal trans-photospheric regions where the sound speed $\sound$
is constant, $\ktr \sim 1/r$ due to the spherical geometry. As
a consequence, waves which are formally propagative at $r=R$ may leak
out to some radius $r > R$ and thence be reflected back towards the
interior, leading to complete wave trapping at frequencies {\em below}
the cut-off $\omegat$.

To address properly this issue of trans-photospheric reflection, it is
necessary to relocate the nominal outer boundary of the star to a
radius at which it is guaranteed that no reflected, inward-propagating
waves will occur. In the context of the linear and adiabatic
approximations adopted herein, this guarantee can only be made of the
outer boundary is located at infinity. However, once non-adiabatic
effects are considered, it is possible that strong radiative or
non-linear dissipation above the photosphere can lead to the effective
absorption of all outward-propagating waves, with no reflection and
subsequent trapping. Such a situation is analogous to the
core-absorption of inwardly-propagating envelope $p$-modes found by
Osaki \shortcite{Osa1977}, and can be treated by placing the outer
boundary at the base of the dissipative region (which may be close to
the photosphere).

Similar arguments concerning the absence of trans-photospheric
reflection can be made for systems with stellar winds. Owocki \&
Rybicki \shortcite{OwoRyb1986} found that, for a line-absorption
driven wind, any wave-like disturbance which reaches the sonic point
$r_{s}$ can never propagate back to smaller radii, due to the
non-linear interaction between the wave and underlying mean flow. In
this case, the outer boundary can be located at $r_{s}$; however,
quantitative treatments are problematical, since the pulsation
equations must be revised to take the sub-sonic wind regions into
account.

In spite of these difficulties, the results of this work are valid on
a phenomenological level, and may be of particular relevance to the 53
Per \cite{SmiKar1976} and slowly-pulsating B (SPB; Waelkens 1991)
classes of variable stars, which are unstable to low-frequency
$g$-mode pulsation due to the metal opacity bump at $\log T \approx
5.3$ (Dziembowski \& Pamyatnykh 1993; Dziembowski, Moskalik \&
Pamyatnykh 1993); the observed periods of these stars are typically
$1-3$ days, which is the same order of magnitude as the trapping
cut-off depicted in figs. \ref{fig:propdiag1} and
\ref{fig:propdiag2}. As indicated previously, the global
self-excitation of a given pulsation mode in one of these stars (or,
indeed, any other type of star) depends on the competitive interplay
between excitation and damping mechanisms, which, respectively, pump
energy into and remove energy from the pulsation at each point within
the star; self-excitation will only occur if the net contributions
from the former outweighs the net deductions from the latter. For the
purposes of the following discussion, both of these generic
energy-transfer processes may be classified as either
\begin{enumerate}
\renewcommand{\theenumi}{(\alph{enumi})}
\item non-adiabatic, where the transfer arises through perturbations to
the specific entropy, corresponding to the operation of a Carnot heat
engine which converts between thermal and mechanical (wave) energy
within a given region,
\label{case:a}
\end{enumerate}
or
\begin{enumerate}
\setcounter{enumi}{1}
\renewcommand{\theenumi}{(\alph{enumi})}
\item advective, where the transfer arises through the non-zero
divergence of the wave flux, corresponding to a net flow of mechanical
energy through the boundaries of the region.\label{case:b}
\end{enumerate}
The opacity mechanism operative in 53 Per and SPB stars is thus a
non-adiabatic excitation mechanism \ref{case:a}, whilst the energy
loss associated with wave leakage at frequencies below the trapping
cut-off $\omegat$ may be identified as an advective damping mechanism
\ref{case:b}.  In the stability calculations of Dziembowski \etal\
\shortcite{Dzi1993}, who use the approach described by Dziembowski
\shortcite{Dzi1977}, the assumption is made that the Lagrangian
pressure perturbation $\delta p$ tends to a limiting value at the
surface; this corresponds to the {\em ab initio} restriction that all
waves are evanescent at the surface, and thus completely trapped
within the star. Hence, any contributions to advective damping arising
from wave leakage are neglected, which {\em might} lead to incorrect
results for the overstability of modes at frequencies below $\omegat$.

However, it must be stressed that the last point is somewhat formal,
if modes are stabilized by non-adiabatic damping well before the
frequency is low enough for leakage to occur; whilst advective damping
might enhance the stability of the virtual modes, it will have little
influence in determining which (trapped) modes are unstable in a
star. In the case of the 53 Per and SPB stars, such a situation may
arise due to the dominance of the opacity mechanism by radiative
damping at lower frequencies. The latter is large for high-order
(large-$n$) $g$-modes, whose eigenfunctions exhibit many radial nodes
in the stellar envelope; the sub-adiabatic temperature gradient in the
radiative parts of the envelope will lead to significant thermal
diffusion between neighbouring oscillating elements, which tends to
suppress the pulsation (see, for instance, equation 26.13 in Unno
\etal\ 1989 plus their accompanying text). These issues will be
examined further in the next paper in this series.

In a contrasting situation, where non-adiabatic damping is less
important at low frequencies, the overstability of virtual modes will
be determined by the relative strengths of non-adiabatic excitation
and advective damping. If the latter is dominant, then $g$-modes will
exhibit an upper limit in their variability periods which corresponds
to the trapping cut-off; in the more rapidly-rotating stars (see, \eg,
Aerts \etal\ 1999), this upper limit will depend, amongst other
things, on $m$ and the degree of rotation. In contrast, if
non-adiabatic excitation dominates, then no upper period limit will be
observed, since virtual modes will be excited in addition to trapped
modes. Estimates of the strength of advective damping can be obtained using,
for instance, the asymptotic approach presented by Shibahashi
\shortcite{Shi1979}. However, as with the local analysis used in
section \ref{sec:dispersion}, such an approach is only valid when the
characteristic variation scale of eigenfunctions is much smaller than
that of the underlying star. This restriction means that Shibahashi's
approach may lead to poor results for those virtual modes with
frequencies close to $\omegat$; therefore, an examination of the
importance of leakage-originated advective damping is deferred to the
following paper, where the pulsation equations are solved globally
using a numerical approach which does not suffer from the restriction
discussed.

Whilst a proper treatment of trapping, even in cases without the
trans-photospheric reflection described above, adds a certain level of
complexity to theoretical studies, it does open the way for
asteroseismological studies of the near-surface regions of early-type
stars. For instance, if an upper period limit is observed as
described, the inferred value of $\omegat$ may be used, in tandem with
equations (\ref{eqn:trapping2}--\ref{eqn:trapping4}), to calculate a
value for the acoustic timescale
\begin{equation}
\tauacc = r/\sound
\end{equation}
in the region where the onset of wave leakage occurs. Since, for an
ideal gas, the adiabatic sound speed $\sound$ is a function of
temperature $T$ alone, this timescale then gives an independent
estimate of the temperature in the outer layers of the star.
Conversely, observations of variability attributable to virtual modes
can confirm the existence of sub-surface convective regions arising in
ionization zones, predicted by evolutionary models of early-type
stars. The degree of wave leakage associated with a virtual mode
depends on the thickness of these regions (which form the
partially-reflective barrier necessary for the existence of virtual
modes); therefore, it might be possible to obtain estimates of the
thickness through measurements of the leakage rate.

\section{Conclusions} \label{sec:conclusions}

The prime conclusion to be drawn from the work presented herein is
that the complete trapping of low-frequency $g$-modes beneath the
surface of early-type stars is not guaranteed. This is especially the
case in rotating stars, where the trapping cut-off frequency $\omegat$
can be significantly increased by the action of the Coriolis force for
all but the prograde sectoral modes. The fact that the latter are more
effectively trapped in rapidly-rotating stars than other types of
modes may explain anecdotal observational evidence which points to
their favoured excitation. As a consequence of the dependence of
$g$-mode eigenfrequencies on the rotation rate, the hypothesis has
been put forward that the set of radial orders $\{n\}$ of trapped
$g$-modes is invariant under the influence of rotation.

Stability analyses which contain the {\em ab initio} assumption of
complete wave reflection at the stellar surface might be in error at
frequencies below the cut-off $\omegat$. More rigorous calculations
can include the possibility of wave leakage, by adopting a more
physically-realistic outer mechanical boundary condition. Such
calculations will reveal to what extent advective damping associated
with leaking virtual modes might suppress the self-excitation of these
modes. These points may be of especial relevance to the 53 Per and SPB
classes of variable stars.

\section*{Acknowledgements} \label{sec:acknowledgements}

I would like to thank Ian Howarth for many useful conversations
regarding wave trapping, and for reading and suggesting improvements
to the manuscript. Also, thanks must go to Conny Aerts and Joris de
Ridder for introducing me to SPB stars. Finally, I am indebted to
Wolfgang Loeffler for the very generous provision of stellar structure
models. All calculations have been performed on an Intel Linux
workstation provided by Sycorax Ltd, and this work has been supported
by the Particle Physics and Astronomy Research Council of the UK.

\bibliography{paper.bbl}

\label{lastpage}

\end{document}